\newcommand{\be}[0]{\begin{equation}}
\newcommand{\ee}[0]{\end{equation}}
\newcommand{\ba}[0]{\begin{eqnarray}}
\newcommand{\ea}[0]{\end{eqnarray}}
\documentstyle[12pt,epsfig]{article}
\begin{document}
\large
\hfill\vbox{\hbox{DTP-96/60}
            \hbox{~July, 1996}}
\nopagebreak

\vspace{2.0cm}
\LARGE
\centerline{\bf Chiral poles and zeros} 
\vspace{0.5cm}
\centerline{\bf and the role of the left hand cut} 
\vspace{0.8cm}
\begin{center}
\large

M. Boglione $^{1,2}$

\vspace{0.5cm}
and
\vspace{0.5cm}

\large

M.R. Pennington $^1$

\vspace{0.6cm}
\normalsize
\begin{em}
$^1$ Centre for Particle Theory, University of Durham\\
     Durham DH1 3LE, U.K.\\

\vspace{0.4cm}

$^2$ Dipartimento di Fisica Teorica, Universit\`a di Torino and \\
      INFN, Sezione di Torino, Via P. Giuria 1, 10125 Torino, Italy\\
\end{em}

\vspace{1.3cm}

\end{center}
\normalsize

\vspace{0.45cm}

{\leftskip = 1.9cm 
 \rightskip = 1.9cm
\centerline {Abstract}
\noindent
It has been recently claimed that the Inverse Amplitude
Method provides a reliable unitarisation of Chiral Perturbation Theory
allowing resonance poles to be accurately uncovered.  We illustrate the
sensitivity of these claims to the treatment of the Adler zero and to 
assumptions about
the left hand cut (and hence about the underlying exchange forces).
Previously favoured methods are shown to mistreat
the Adler zeros and violate crossing symmetry casting doubt on the 
precision of their
phenomenology. A more reliable solution is proposed.
\par}
\newpage
\baselineskip=7mm
\parskip=3mm
\noindent {\bf {\large 1. Introduction}}
\vspace{0.3cm} 

\noindent
Developments in chiral perturbation theory ($\chi$PT) have led to a
resurgence of interest in low energy $\pi \pi$ scattering~\cite{GL,DonHol}.
The predictions
of $\chi$PT unambiguously apply to the $\pi \pi$ process in the subthreshold
region. Chiral dynamics demands that $\pi \pi$ and $\pi K$ amplitudes have
zeros below threshold. These are the on-shell appearance of the Adler zero
and they occur in each $S$-wave amplitude~\cite{Weinberg}.
These zeros of course occur in physical amplitudes. 

\noindent
Recently, much
attention has been paid to how to compare the predictions of $\chi$PT
with experiment~\cite{MIT,Trento}. A key discussion point concerns how far
one can reliably
continue the $\chi$PT amplitudes above threshold to where experimental
information exists? It is a feature of the physical world that resonances
dominate the behaviour of isospin $I\leq 1$ partial waves. Resonances, however,
do not appear in $\chi$PT at any finite order, since low energy resonant
amplitudes are non-perturbative in their fulfillment of unitarity.
Consequently, the issue of how to extract resonance physics from the chiral
expansion has been a topic of heated   debate~\cite{MIT,Trento}.

\noindent Dispersion relations provide an invaluable connection between 
scattering
amplitudes in one energy region and another. They connect the subthreshold
region where $\chi$PT applies and the world of resonance physics.
However, their evaluation requires knowledge not just of the singularity
structure of the amplitude, but of the exact form of its discontinuity
across any cut; for example we must know the imaginary part to determine
the real part of each amplitude. A seemingly significant advantage of 
considering the
dispersive representation of the inverse of a partial wave amplitude is
that its right hand cut discontinuity is just given by phase space in the
elastic region, thanks to unitarity.
Consequently, when the integral along this cut is controlled by the low
energy region, it can be evaluated reliably without any further
information. While what we might regard as \lq\lq kinematics"
fixes the right hand cut, dynamics is built in by assumptions
about the left hand cut. Indeed, this is the basis of the $N/D$
  method~\cite{Eden}. 

\noindent 
While partial wave amplitude, $t(s)$, have right and left hand cuts,
unitarity does not  allow them to have poles on the physical sheet. As a
consequence, their inverses, $f(s)$, have similar right and left cuts.
However, partial wave amplitudes can and do have zeros, which in turn means
their inverses have poles on the physical sheets, making their singularity
structure a little more  complicated. These poles are the essence of chiral
  dynamics. 

\noindent
Very recently Dobado and Pel\'aez~\cite{DP96} have used 
dispersion relations for the
inverse $\pi \pi$ and $\pi K$ amplitudes and ignored the appearance of such
poles. This has allowed them to {\it derive} a Pad\'e-like approximant as
the sum
of chiral perturbative predictions for the $S$-wave amplitudes, rather like
that much used for the $P$-wave~\cite{Truong}.  Their choice of
approximations gives a
description that does agree with experiment. However, we show here
how strongly these essentially dispersive continuations
of the predictions of $\chi$PT depend on
\begin{description}
{\leftskip = 2cm 
 \rightskip = 2cm
\item (i) the left hand cut discontinuity,
\item (ii) the existence of chiral poles,
\item (iii) additional summation assumptions implicit in the 
Pad\'e approximation.
\par}
\end{description}
Indeed, within each set of assumptions, good agreement with
experiment below 800 MeV is possible, depending on different choices
of the ${\cal O} (p^4)$  $\chi$PT parameters ${\overline \ell_i}$ 
for $i=1,2$~\cite{GL}.
Thus this is not a reliable way of determining the  
${\overline \ell_i}$. 

\noindent
In the next section we give a general introduction to an inverse
amplitude method and the treatment of chiral poles.  We propose
several alternative forms for the left hand cut discontinuity and in sect.~3
show how the physical region results depend on these. In Sect.~4 we
give our brief  conclusions.  
\vspace{8mm}

\noindent {\bf {\large 2. The amplitude and its inverse}} 

\noindent
Consider the $\pi \pi$ partial wave amplitudes, $t^I_J(s)$, with isospin $I$
and spin $J$. Defining $s$ to be usual square of the c.m. energy and denoting
the pion mass by $\mu$, elastic unitarity requires
\be
{\rm Im }\;t^I_J(s) = \rho (s) \; | \, t^I_J(s) \, |^{\, 2}
\ee
where $\rho (s) \equiv \sqrt{1-4 \mu ^2/s}$.
In practice, the first significant inelastic channel opens up only at $K
\overline K$ threshold. $J \le 1$ partial waves control $\pi \pi$
scattering in this region. Now each $S$ and $P$-wave, $t(s)$ (dropping $I,\, J$
labels for simplicity), has a zero at $s=s_0$. For the $S$-waves, 
these are
demanded by the Adler condition; for the $P$-wave, the zero is kinematic
being at threshold so $s_0 = 4\mu ^2$. \footnote{Higher partial waves have
higher order zeros, of course.} The inverse, $f(s)$, of each of these
partial wave amplitudes will thus have a simple pole at $s=s_0$, with
residues $r$. Consequently we define
\be
f_{pole} (s) \equiv \frac{r}{s-s_0}\,.
\ee
Now we shall assume that $|\, f(s) \, |<s$ as $s \to \infty$, so that each 
 $f(s)$  satisfies a once subtracted dispersion relation with $s=s_{1}$,
the \vspace{3mm} subtraction point,
\ba
f(s) = f_{pole}(s) + c & + & \frac{(s-s_1)}{\pi} \, 
\int _{4\mu ^2} ^{\infty} 
\frac{ds'}{(s'-s_1)(s'-s)} \; {\rm Im } \, f(s') \, \nonumber \\ & + & \, 
\frac{(s-s_1)}{\pi} \, \int _{-\infty} ^{0} \frac{ds'}{(s'-s_1)(s'-s)} \; 
{\rm Im } \, f(s')\quad .    \vspace{5mm}
\ea 

\noindent ${\rm Im } \, f(s)$ is to be evaluated above both cuts; 
$c$ is a constant 
simply related to the value
of the inverse amplitude at the subtraction point $s=s_1$.
Care must be taken if $s_1$ is chosen to be the position
of the simple pole in Eq. (2), i.e. $s_1=s_0$, then
$$ c\,=\,\lim_{s\to s_0}\; (f(s) -f_{pole}(s))\quad .$$
Since $\chi$PT is believed to describe $\pi\pi$ scattering
in the subthreshold region quite accurately, the idea is to use
its predictions to fix the position of the chiral pole
($s=s_0$) in the inverse amplitude and its residue $r$
and the subtraction constant $c$, from the value of $f(s)$
at the subthreshold subtraction point. The partial wave may well converge
faster than we assume Eq.~(3), so that subtractions are not needed. 
However, the
advantage of making a subtraction is that the more distant parts of the cut
discontinuities play less of a role for the region of interest, $-0.5 <
(s - 4 \mu ^2) < 0.5$ (GeV$^2$). The physics contained in these distant and
poorly known contributions is replaced by the subthreshold subtraction term
fixed by $\chi PT$. 

\noindent 
To evaluate the dispersive representation of Eq.~(3), we must know
the imaginary parts; clearly  
\be
{\rm Im } \, f(s) = {\rm Im } \, \frac{1}{t(s)} = 
- \; \frac{ {\rm Im } \, t(s)}{|\, t(s) \,
|^{\, 2}}\quad .
\ee 
For $4 \mu ^2 < s \le 1$ GeV$^2$,
this just equals $- \rho$ in the elastic region, Eq.~(1). 
The fact that the low energy right hand cut is specified by
non-linear unitarity, Eq.~(1), is a key step in building non-perturbative
(and hence resonance) physics into the amplitudes $t(s)$. Indeed, since the
elastic region extends essentially up to $K \overline K$ threshold, the
right hand cut contribution of Eq.~(3) can be reliably computed up to 750 MeV
or so for the $I=0$ $S$-wave and even higher for the  $I=2$ $S$-wave
and $I=1$ $P$-wave, for which the
inelasticity is naturally   weaker. We  therefore set
${\rm Im}\, f(s)\, = \, - \rho(s)$ everywhere along the right
hand cut as a reasonable approximation.  The right hand cut integral
of Eq.~(3) can then be
performed analytically and involves the Chew-Mandelstam function
${\overline J(s)}$ of one loop
$\chi$PT~\cite{GL},  so that the integral is just 
$-16 \pi ( {\overline J(s)}\,-\,
{\overline J(s_1)})$. However, the left hand cut is also crucial
and we adopt a number of different schemes for its calculation,
 which we now   describe. 

\noindent 
{\bf Scheme~I} involves the simplest of all assumptions and 
sets the left hand cut discontinuity equal to zero.  Of course,
this violates crossing symmetry.  It is impossible for the right hand cut
to be non-zero and the left hand cut, generated by exchange forces, to be zero
for amplitudes that are crossing symmetric as $\pi\pi$ is. Nevertheless,
this provides a base from which to judge the dependence on the assumed left
hand cut. The dispersive representation
of Eq.~(3) is evaluated with the subtraction constant specified
by one loop $\chi$PT at $s_1\,=\,4\mu^2/3$ ---
 a point conveniently between the two cuts.
This we call   scheme~I. 

\noindent
{ \bf Scheme~II} is to assume that the left hand cut discontinuity
is given by one loop $\chi$PT out to $s=-(M^2 -4\mu^2)$ with
$M$ typically 0.5-0.6 GeV and to assume it to be constant
beyond that.  This is in keeping with the notion that
$\chi$PT predictions may be believed for $t,u \leq$ (0.6 GeV)$^2$
and that the distant left hand cut should not have too big an effect on the 
direct channel
amplitude for $s\,<\,1$   GeV$^2$. The subtraction term is fixed as in 
Scheme~I
with $s_1\,=\,4\mu^2/3$. As an illustration we plot in Fig.~1 the inverse 
$I=0$ $S$-wave 
amplitude (as an example) in Scheme II. We see how the subthreshold pole 
dominates the low
energy amplitude throughout the region of $|s| < 1$ GeV$^2$. 
This spotlights the
perils of neglecting its appearance~\cite{DP96}.  In no sense
can the residue of the pole be regarded as small~\cite{DP92}.
In Fig.~2 we plot  
$f(s)-f_{pole}(s)$, again for the $I=0$ $S$-wave, to indicate the
difference using Scheme I or II makes to the physical region amplitude
we wish to compare with experiment.

\noindent 
{ \bf Scheme~III} is a stronger assumption, closer to that of 
Dobado and Pel\'aez~\cite{DP96}.
This is to note that one loop $\chi$PT satisfies unitarity
(perturbatively) along the right hand cut, viz. Eq. (1),
by
\be
{\rm Im \,} t^{(1)} (s)\,=\,\rho \,\mid t^{(0)}(s)\mid^2,
\ee
where the bracketed superscripts label the order in the
chiral expansion at which the whole partial wave is computed.
It is then useful to note that the tree level amplitudes
have the simple structure
\be
t^{(0)}\,=\,{{s-s_0^{(0)}}\over{r^{(0)}}}
\ee
where $s_0^{(0)}$ is the position of the zero
(Adler zero for the $S$-waves, threshold zero for the $P$-wave)
and $r^{(0)}$ is the residue of the corresponding pole in the
inverse amplitude. As is well-known~\cite{Weinberg},
\ba
s^{(0)}_0\; & = & \; \mu^2/2\quad\;\;{\rm for\;the\;}I=0\;\,S{\rm -wave},
\nonumber \\
            & = & \; 4\mu^2\qquad{\rm for\;the\;}I=1\;\,P{\rm -wave},
\\
            & = & \; 2\mu^2\qquad{\rm for\;the\;}I=2\;\,S{\rm -wave} .
\nonumber 
\ea
with residues
\ba
r^{(0)}_0\; & = & \; 16\pi\, F^2\qquad{\rm for\;the\;}I=0\;\,S{\rm -wave},
\nonumber \\
            & = & \; 96\pi\, F^2\qquad{\rm for\;the\;}I=1\;\,P{\rm -wave},
\\
            & = & \! -32\pi\, F^2\quad\,\;\; {\rm for\;the\;}I=2\;
\,S{\rm -wave} .\nonumber 
\ea
where $F$ is the pion decay constant in the chiral limit, 
$F = 0.94 F_{\pi}$ viz.~\cite{GL}.
In scheme~III we assume along the left hand cut too that
\be
{\rm Im \, } \frac {1}{t(s)}\,\equiv \,- \frac{{\rm Im} 
t(s)}{\mid t(s)\mid^2}\,
\simeq\,\, -\frac{{\rm Im} t^{(1)}(s)}{(t^{(0)}(s))^2} \, \; .
\ee
This allows us to avoid explicitly evaluating a dispersive
representation~\cite{Truong,DP96}, as follows~: the one loop $\chi$PT 
amplitude
$t^{(1)}$ grows asymptotically like $s^2$ modulo logarithms.
Let us consider \vspace{3mm} the function
\ba
\Delta t(s)\,\equiv\,t^{(1)}(s)\,&-&\,t^{(1)}(s_1)\,-\, (s-s_1)\,
\frac{d}{ds} \, t^{(1)}(s_1) \nonumber \\
&-&\,\frac{1}{2} (s-s_1)^2\,
\frac{d^2}{ds^2} \, t^{(1)}(s_1) \vspace{6mm}  
\ea

\noindent where $s_1$ is again a subtraction point in the subthreshold 
region.
Then the function $\Delta t(s)/(s-s_1)^2$ will satisfy a once subtracted
dispersion relation with zero subtraction constant, \vspace{3mm} so that
\ba
\! \! \! \! \frac{ \Delta t(s)}{(s-s_1)^2} &=&  \frac{(s-s_1)}{\pi} \, 
\int _{4\mu ^2} ^{\infty} \ ds'  
\frac{{\rm Im } \, t^{(1)}(s')}{(s'-s_1)^3 (s'-s)} \; \nonumber \\
& & \;\; + \; \frac{(s-s_1)}{\pi} \, \int _{-\infty} ^{0} \ ds' 
\frac{{\rm Im } \, t^{(1)}(s') }{(s'-s_1)^3 (s'-s)} \; \vspace{6mm} 
\ea

\noindent This is just the statement that the amplitudes of $\chi$PT
are analytic in the cut plane.
Conveniently, choosing the subtraction point $s_1$ to be the position of
the tree level zero, i.e. $s_1=s_0^{(0)}$, \vspace{3mm} 
we have noting Eq.~(6)
\ba
\! \! \! \! \frac{ \Delta t(s)}{(t^{(0)}(s))^2} &=& 
\frac{\left(s-s_0^{(0)}\right)}{\pi} \, 
\int _{4\mu ^2} ^{\infty} \ ds'
\frac{{\rm Im } \, t^{(1)}(s)}{\left(s'-s_0^{(0)}\right)
(s'-s)(t^{(0)}(s'))^2} \;
\nonumber \\ 
& & \;\; + \; \frac{\left(s-s_0^{(0)}\right)}{\pi} \, 
\int _{-\infty} ^{0} \ ds' 
\frac{{\rm Im } \, t^{(1)}(s)}{\left(s'-s_0^{(0)}\right)
(s'-s)(t^{(0)}(s'))^2} \; \vspace{6mm} 
\ea

\noindent Now using Eqs.~(5,9),
Eq.~(3) and Eq.~(12) can be simply \vspace{3mm} added to give :
\ba
\frac{1}{t(s)} &=& \frac{r}{s-s_0} \,+ \, c \,- \,
\frac{t^{(1)}(s)}{(t^{(0)}(s))^2} + 
\frac{t^{(1)}(s_0^{(0)})}{(t^{(0)}(s))^2} \nonumber \\
&+& \frac{r_0}{t^{(0)}(s)} \; \frac{d\,}{ds}\,t^{(1)}(s_0^{(0)})\, 
+\, \frac{1}{2}
(r_0)^2 \; \frac{d^2\ }{ds^2}\,t^{(1)}(s_0^{(0)}) \vspace{6mm}
\ea

\noindent Uniquely for the $P$-wave, the zero of the full
partial wave amplitude and that of the tree level approximation
are at the same position, viz. $s^{(0)}_0\,=\,s_0\,=\,4\mu^2$. 
Then for the
$P$-wave amplitude, \vspace{3mm} we have
\ba
\frac{1}{t(s)} \,=\, \frac{1}{t^{(0)}(s)} \, 
\left( \frac{r}{r_0} + \frac{r_0}{r} \right) 
&-& \frac{t^{(1)}(s)}{(t^{(0)}(s))^2}  \nonumber \\ &-& 
\frac{1}{2} \, (r^2-r_0^2) \, \frac{d^2}{ds^2} \, t^{(1)}(s_0^{(0)}) 
\vspace{6mm}
\ea

\noindent If we further assume that the $P$-wave scattering length
for the full $\chi$PT amplitude is the same as for the
tree level approximation, then $r=r_0$ and we obtain
\footnote{n.b. $t^{(1)}(s)$ means the full one loop
partial wave and not just the one loop correction of 
Refs.~\cite{Truong,DP96}.}
$${1\over t(s)}\;=\;{2\over t^{(0)}(s)}\,-\,{t^{(1)}(s)
\over(t^{(0)}(s))^2}$$
i.e.
\be
t(s)\;=\;{{t^{(0)}(s)^2}\over{2 t^{(0)}(s) - t^{(1)}(s)}}
\ee
which is the $[1,1]$ Pad\'e approximant
introduced by Truong~\cite{Truong}. But note the key chain of 
assumptions needed to
deduce   this. 

\noindent
This brings us to {\bf scheme~IV}. This is again to
assume the left hand cut discontinuity for all $s < 0$ is given by Eq.~(9)
and further to assume for each all orders partial wave amplitude
that the position of the zero and the slope there are both just as
for the tree level amplitude, i.e. $s_0 = s_0^{(0)}$
and $r = r_0$. This gives Eq.~(15) for each $S$ and $P$-wave amplitude.
This is the scheme~of Dobado and Pel\'aez~\cite{DP96}.
Of course, the $S$-wave zeros crucially move making this approximation
poor, as we shall   see. 

\noindent 
Each of these schemes provides a continuation into the physical region of
the predictions of $\chi$PT, that are assumed exact in
the neighbourhood of the subtraction point in
the subthreshold region.
Though of course the underlying chiral amplitudes
satisfy crossing symmetry exactly at each order, these
continuations do not.  We can test this failure
by evaluating the five crossing sum rules that involve
just the $S$ and $P$-waves.  If these relations were
exactly satisfied they guarantee that there exists
a $\pi\pi$ amplitude with the correct crossing properties
with these precise $\ell=0$ and 1 partial waves.
In the appendix, we detail what these relations are
and specify a measure by which we tell
how well each of our approximations satisfies crossing.
We tabulate the results in the next  section. 

\vspace{8mm}

\noindent {\bf {\large 3. Results}}
\vspace{0.3cm} 

\noindent In this section we display the results of the calculation 
of the 3 
lowest partial wave amplitudes according to each of the four Schemes 
(described
in the last section) for approximating the left hand cut contribution.
To determine the subtraction term $c$ and pole position $s_0$ and 
residue $r$
of Eqs.~(2,3) from $\chi$PT to one loop order, we have to choose 
values for the
parameters ${\overline \ell_i}$ of the $SU(2)$ Chiral Lagrangian.
As a guide we take the values~\cite{PP}
\be
\overline \ell_1 = -0.3, \ \; \overline \ell_2= 4.5\; . 
\ee
\noindent
The formulae for the $\pi\pi$ invariant amplitude to one loop are taken from
Eq.~(17.1) of Ref.~\cite{GL}, in which recall
$F = 0.94 F_{\pi}$ where $F_{\pi} = 93$ MeV.  
The resulting elastic partial waves can then be expressed in terms of the
corresponding phase-shift $\delta ^I_J$ by way of the standard representation
\be
t^I_J(s)\;=\;\frac{1}{\rho}\,\sin \delta^I_J\, \exp \left( i \delta
^I_J \right) \qquad.
\ee 
In Figs.~3-5, we show the phase-shifts for the $I=0, 2$ $S$-waves 
and the $I=1$
$P$-wave from each of the calculational schemes, together with 
experimental data
from the LBL analysis of Protopopescu et al.~\cite{LBL}, the 
CERN-Munich results of
Ochs~\cite{Ochs} and of Hoogland et al.~\cite{Hoog},
and the $K_{e4}$ results of Rosselet et al.~\cite{Ke4}. 
The curves I, II, III
show rather dramatically how changing the left hand cut discontinuity, 
while keeping the same underlying chiral perturbative amplitude
(i.e. the same subtraction constant $c$, pole position $s_0$ and 
residue  $r$
in Eqs.~(2,3,13)), alters the partial waves in the physical region.  
Fig.~4 illustrates
how the left hand cut (which is produced by exchange forces) determines the
generation of the $\rho$-resonance --- a fact on which the {\it bootstrap}
principle was based~\cite{Eden}. 
Changing from curves III to IV illustrates the effect of assuming
the chiral pole's position and residue are as in the tree level amplitude
(without altering the left hand cut discontinuity),
implicit in the Pad\'e-like summation of Refs.~\cite{Truong,DP96}.
This demonstrates that a scheme like IV cannot be regarded as  
an accurate way of determining the Lagrangian parameters ${\overline \ell_I}$
$(i=1,2)$. In fact, we see that, with the choice of the
${\overline \ell_i}$ of Eq.~(16), Scheme~II provides the best agreement with
data.
However, by a suitably different choice of the Lagrangian parameters
any of the other Schemes can be made to agree better with experiment ---
but not for all of the waves at one time as we now explain.

\noindent An independent way to test the consistency of each 
approximation scheme
is to check how well crossing symmetry is fulfilled by the
resulting 3 lowest partial
waves.
The problem of how to express the consequences of crossing symmetry ---
a property of the full amplitude --- in terms of a finite
number of  partial waves
was solved more than 25 years ago by Balachandran and Nuyts~\cite{BN} 
by considering the amplitudes in the Mandelstam triangle.  An explicit
realisation of these subthreshold relations, known as 
{\it crossing sum rules},
was given shortly thereafter by Roskies~\cite{RoskCP}. 
These provide a necessary and sufficient
set of conditions for crossing.  In the Appendix we give the five sum 
rules that involve just the $S$ and $P$-waves. 
\begin{table}[h]
\begin{center}
\begin{tabular}{|l|c||c|c|c|c|c|} 
\hline
\rule{0cm}{0.7cm} Scheme&l.h.cut &cross1&cross2&cross3&cross4&cross5\\
\hline \hline
\rule{0cm}{0.7cm} I     & none   &  1.0 & 0.6  &  0.5 &  0.9 &  0.6  
\\ \hline
\rule{0cm}{0.7cm} II    &Eq.~(3) &  0.2 & 0.1  &  0.1 &  0.2 &  0.0  
\\ \hline
\rule{0cm}{0.7cm} III   &Eq.~(13)&  1.0 & 0.2  &  1.1 &  1.3 &  0.9  
\\ \hline
\rule{0cm}{0.7cm} IV    &Eq.~(15)&  1.3 & 0.0  &  1.5 &  1.5 &  1.2  
\\ \hline
\end{tabular}
\caption{\leftskip = 2cm 
 \rightskip = 2cm{Tests of the crossing sum rules, Eqs. (A1-5), 
as defined by the ratio $R$ in Eq.~(A6) each expressed as a percentage.}}
\end{center}
\end{table}

\noindent In Table I we show how the
partial waves calculated in each scheme fulfil these relations in terms
of the measure defined in Eq.~(A6). We see that if the partial waves have no
left hand cut (Scheme~I) crossing is violated by 0.5-1.0\%. In contrast, 
if the nearby part of the left and right hand cuts are given by
one loop $\chi$PT (Scheme~II), then the violation is only 0.1-0.2\%.  
Since having no left hand cut,
but explicitly having a right hand cut, is clearly in violation of crossing 
symmetry, the 1\% violation of Scheme~I sets the scale for the level of 
violation.
As already mentioned, by making the underlying chiral amplitude different by
a different choice of the ${\overline \ell_i}$ a partial wave in any
scheme can be brought into agreement with experiment. However, only for
Scheme~II with its near crossing symmetry can this be achieved for each
partial wave simultaneously. 
We see that Schemes~III and IV, in which the left hand cut for the inverse 
amplitude is approximated by Eq.~(9), give a larger violation than even 
having no left hand cut. 

\vspace{8mm}

\noindent {\bf {\large 4. Conclusions}}
\vspace{0.3cm} 

\noindent Dispersion relations for the inverse partial wave amplitudes
provide a method of imposing a right hand cut structure consistent with 
unitarity.
Thus this method is a useful way of continuing the predictions of $\chi$PT 
at any order,
into the physical regions where the non-linearity of  unitarity
determines resonant behaviour.
 Considering $\pi \pi$ scattering, we have shown here how strongly 
these continuations depend on the assumed left hand cut discontinuity
and on our knowledge of the position and residue
of the subthreshold poles 
that are the key embodiment of chiral dynamics. We have shown how 
crossing symmetry allows us to select
 between different approximation schemes.  Not surprisingly
 neglecting the left hand
 cut (and by inference exchange forces) violates crossing.  However we 
 have seen that the  favoured Pad\'e-like
  sums violate crossing even more strongly.  Consequently, calculations 
based on such approximation
  schemes cannot be regarded as reliable ways of determining the
  Chiral Lagrangian parameters ${\overline \ell_1},\,{\overline \ell_2}$.
  
\noindent The inverse amplitude method is a way of unitarising the
predictions of $\chi$PT.  However, different assumptions on how to 
implement the method
have a considerable effect on the physical region predictions and the
subsequent 
comparison with data.
  Reassuringly, the requirement of crossing symmetry brings closer 
agreement with experiment
  as our Scheme~II demonstrates. This suggests a more reliable 
continuation of $\chi$PT
  into the physical region could be obtained by using  the crossing 
sum rules to restrain
  the form of the left hand cut discontinuity and the corresponding 
values of the Lagrangian
  parameters ${\overline \ell_1},\,{\overline \ell_2}$.  Then the 
inverse amplitude
  method might achieve the precision phenomenology earlier 
treatments claim and be
  able to predict the resonance poles that control low energy meson 
scattering processes.
\newpage
\begin{center}
{\bf Acknowledgements}
\end{center}
One of us (MRP) is grateful to
Dominique Toublan, Res Urech, Fernando Cornet and Jorge Portol\'es
for initial discussions about the Inverse Amplitude Method.
This would not have been possible without (i) the
organisation of the \lq\lq Workshop on the Standard Model at
Low Energies" by Hans Bijnens and Ulf Meissner that took
place at ECT* in Trento in May 1996 and (ii) the support of the
EURODA$\Phi$NE Network that provided funds to attend this meeting
under grant ERBCMRXCT920026 of the EC Human
and Capital Mobility programme.
To both ECT* and the EU thanks are due.
The other  author (MB) is grateful to M. Anselmino, E. Predazzi 
and R. Garfagnini for
permitting her to study at the University of Durham and to INFN 
(Sezione di
Torino) for the necessary travel support.
\newpage
\setcounter{equation}{0}
\renewcommand{\theequation}{A\arabic{equation}}
\begin{center}
{\bf Appendix A}
\end{center}

\noindent Below we give the integral relations  crossing symmetry imposes
on the $\pi\pi$ partial wave amplitudes, $t^I_J(s)$, with isospin $I$
 and spin $J \leq 1$~\cite{RoskCP}~:
\be
\int _0 ^{4\mu ^2} ds \, (4\mu ^2 -s)\, (3s-4\mu ^2)\,
\left( t^0 _0 (s) + 2 t^2 _0(s)\,\right) = 0
\ee
\be
\int _0 ^{4\mu ^2} ds \, (4\mu ^2 -s)\,
\left( 2 \ t^0 _0 (s) - 5 t^2 _0(s)\, \right) = 0
\ee
\ba
& & \int _0 ^{4\mu ^2} ds \, (4\mu ^2 -s)\, (3s-4\mu ^2)\,
\left(2 \ t^0 _0 (s) - 5 t^2 _0(s)\, \right) \nonumber \\
& & \hspace{4.3cm} + 9 \int _0 ^{4\mu ^2} ds \, 
(4\mu ^2 -s)^2  \, t ^1 _1 (s) = 0 
\ea
\ba
& & \int _0 ^{4\mu ^2} ds \, (4\mu ^2 -s) \, s^2 \, 
\left(2 \ t^0 _0 (s) - 5 t^2 _0(s)\,\right) \nonumber \\
& & \hspace{4.3cm} + 3 \int _0 ^{4\mu ^2} ds 
\,(4\mu ^2 -s)^3 \, t ^1 _1(s) = 0 
\ea
\ba
& & \int _0 ^{4\mu ^2} ds \, (4\mu ^2 -s)^2  \, s^2 \, 
\left(2 \ t^0 _0 (s) - 5t^2 _0(s)\,\right) \nonumber \\
& & \hspace{3cm} + 3 \int _0 ^{4\mu ^2} ds \,(4\mu ^2 -s)^2 \, 
(8\mu ^2 -3s) \, s \, t ^1 _1 (s) = 0 \qquad .
\ea
Each of these  relations can be written generically as
$$
\int _0 ^{4\mu ^2} ds \,\omega(s)\, \sum_I \alpha_I\ t^I_J(s)\;=\;0\quad .
$$
where the $\alpha_I$ are constants.
A measure of how close any integral is to zero can be assessed by
forming the ratio
\be
R\;=\;\frac {\int _0 ^{4\mu ^2} ds \,\omega(s)\, \sum_I \alpha_I\ t^I_J(s)}
{\int _0 ^{4\mu ^2} ds \,\omega(s)\,\mid \sum_I \alpha_I\ t^I_J(s)\mid}
\ee
This is the quantity expressed as a percentage that we quote in Table I
for each of the five sum rules of Eqs.~(A1-5). As is well known the 
tree level
amplitudes~\cite{Weinberg} satisfy crossing exactly, we could have 
defined an alternative measure 
by replacing each partial wave $t^I_J(s)$ by its difference from the
tree level approximation. This gives values for the corresponding ratio $R$
a factor of two larger for all the numbers in Table I, leaving the qualitative 
comparison the same.

\newpage
\textwidth 18cm
\textheight 25.5cm
\topmargin -3cm
\begin{figure}[h]
\vspace{4cm}
\begin{center}
\mbox{~\epsfig{file=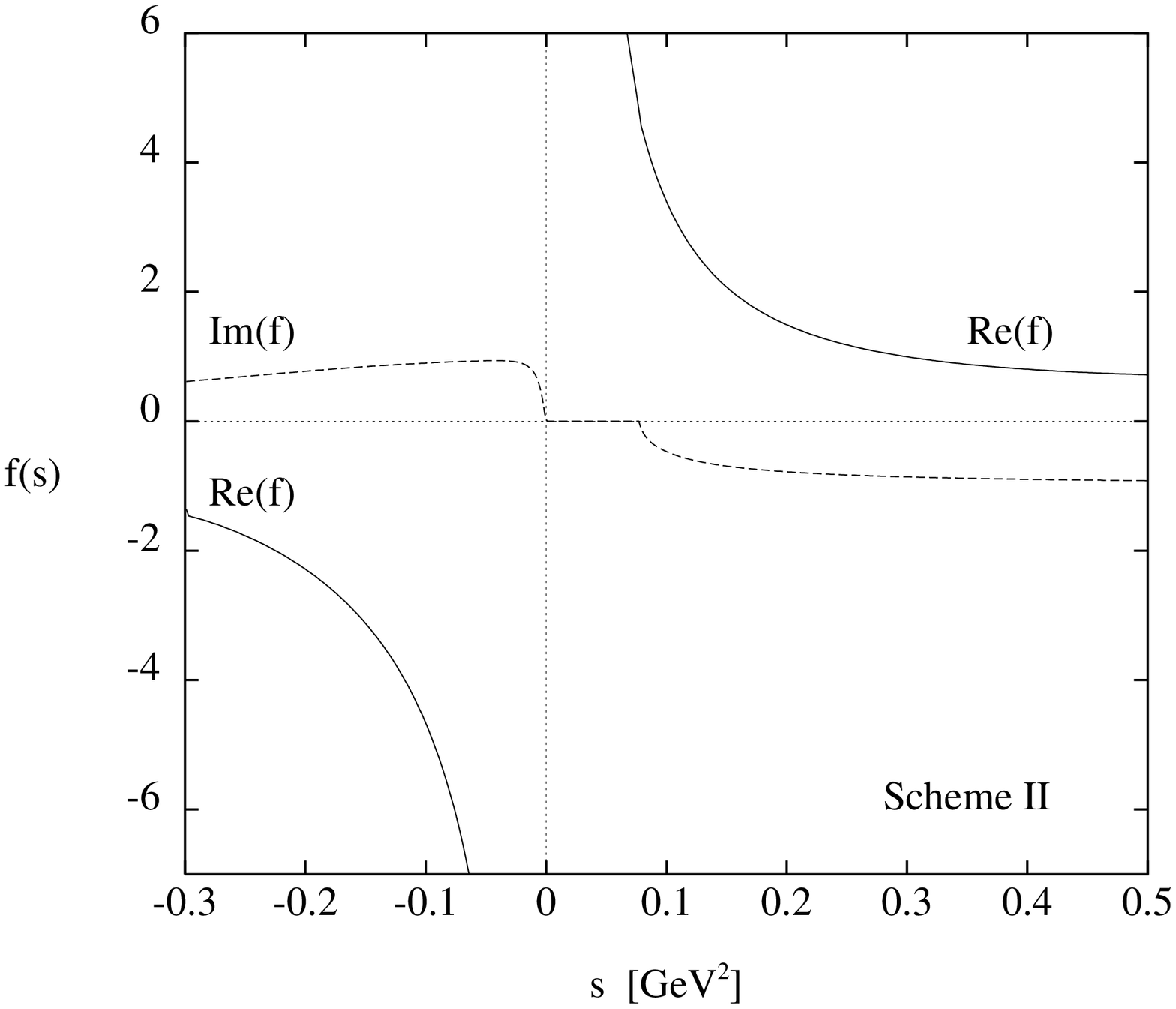,angle=-90,width=15cm}}
\caption{\leftskip = 2cm 
Real and imaginary parts of the $\pi \pi$ $I=0$ $S$-wave inverse amplitude,
$f(s)=1/t^0_0(s)$ for $s  + i \epsilon$, from Eq.~(3), 
calculated using Scheme II.  Note the way the pole, which is the Adler zero
in the partial wave, dominates the behaviour of the inverse.}
\end{center}
\end{figure}
\begin{figure}[t]
\begin{center}
\mbox{~\epsfig{file=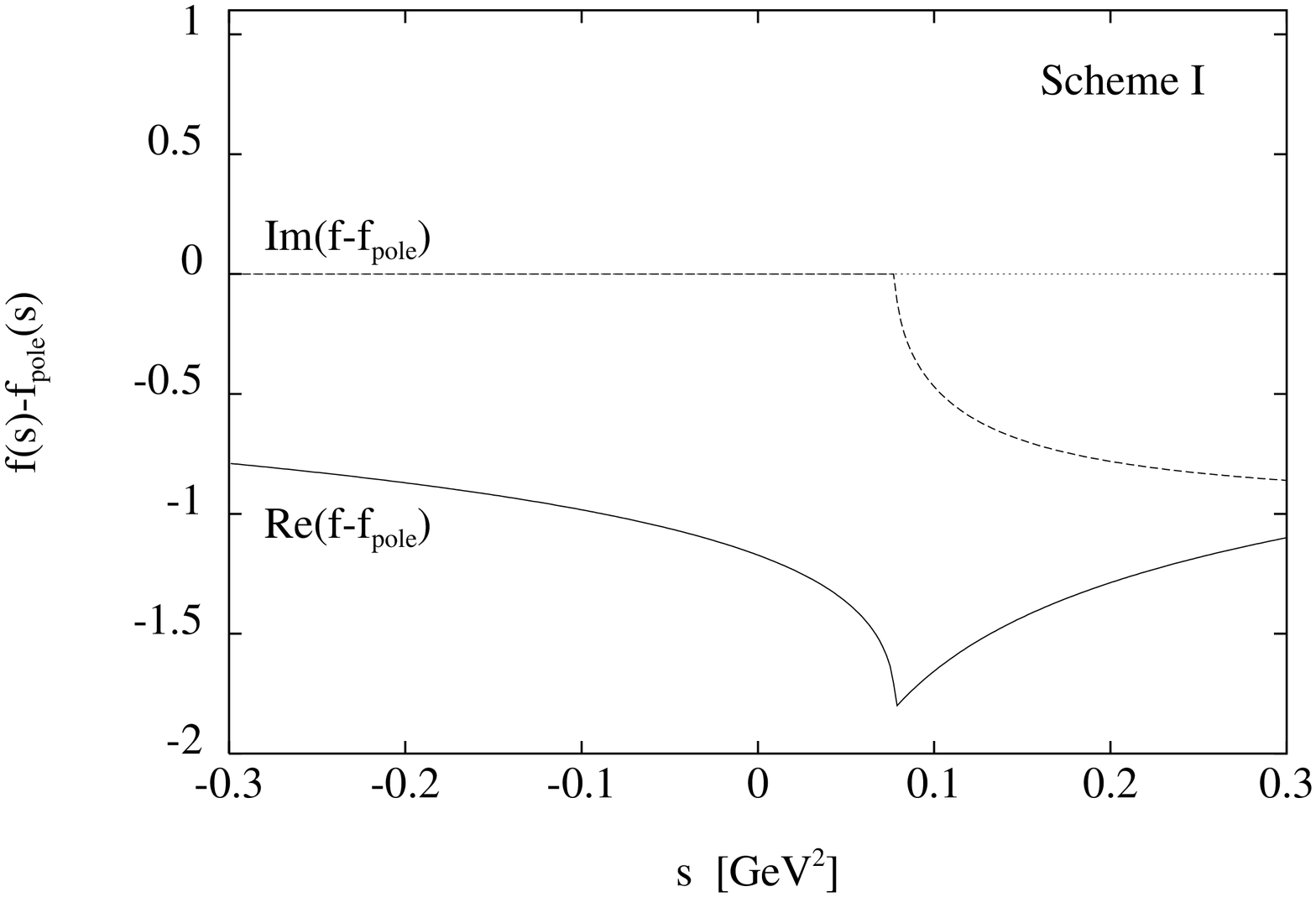,angle=-90,width=14cm}}
\end{center}
\end{figure}
\begin{figure}[b]
\vspace{-5cm}
\begin{center}
\mbox{~\epsfig{file=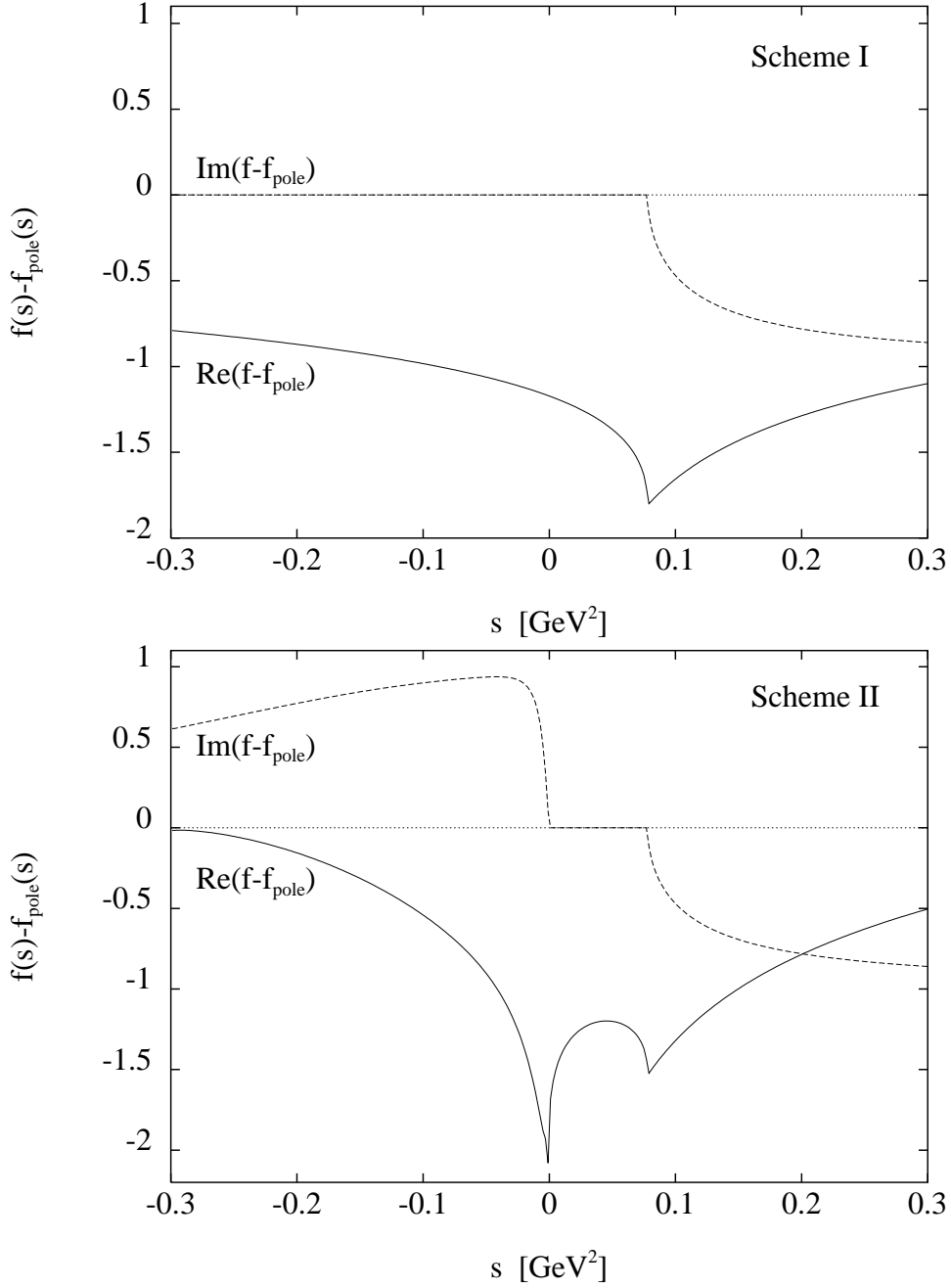,angle=-90,width=14cm}}
\caption{\leftskip = 2cm 
Real and imaginary parts of the $\pi \pi$ $I=0$ $S$-wave inverse amplitude
without the contribution of the pole,
$f(s)-f_{pole}(s)$, for $s  + i \epsilon$, calculated from Eq.~(3)
according to  Scheme~I, which has no left hand cut, and
Scheme~II. Notice how the behaviour of the real
part (solid line) of the amplitude at $s=0$ and $s=4\mu^2$ reflects the
 strength of the
relevant cut discontinuity, which is just the imaginary part (dashed line).}
\end{center}
\end{figure}
\begin{figure}[p]
\begin{center}
\mbox{~\epsfig{file=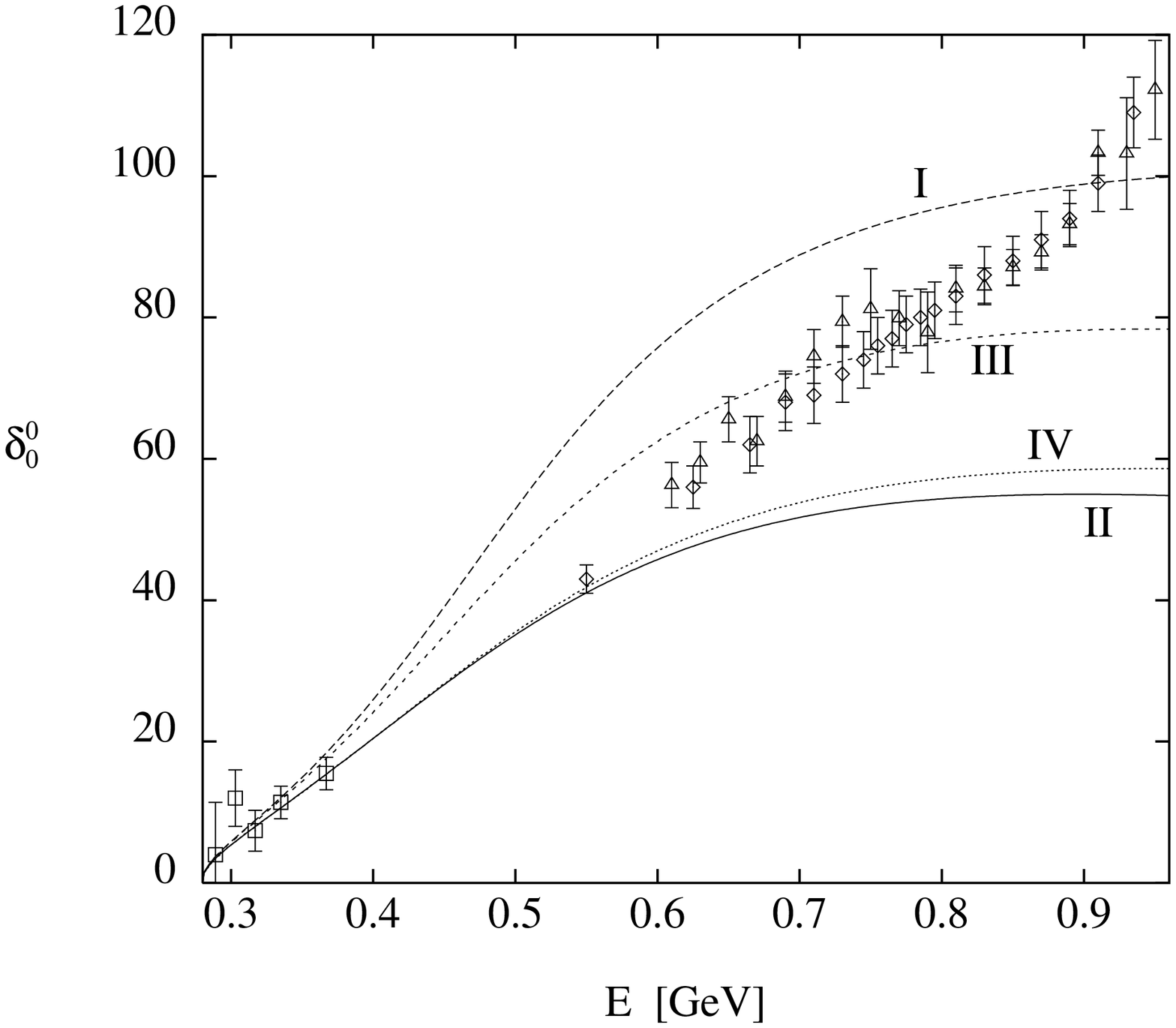,angle=-90,width=15cm}}
\caption{\leftskip = 2cm 
The $\pi \pi$ $I=0$ $S$-wave phase shift, $\delta ^0 _0$ in degrees, below $K
\overline K$ threshold as a function of $\pi\pi$ mass 
$E\,=\, \protect \sqrt{s}$.
The dashed line  is the result
obtained with no left hand cut contribution (Scheme~I); the solid 
line comes from the explicit evaluation of the left hand cut dispersive 
integral (Scheme~II); the dotted lines are obtained with additional
summation assumptions (Schemes~III and IV). The experimental results are from  
Protopopescu et al.~[10] (diamonds), the energy-independent 
analysis by Ochs~[11] (triangles) and the $K_{e4}$ decay data of 
Rosselet et al.~[13] (squares).}
\end{center}
\end{figure}
\newpage
\begin{figure}[p]
\begin{center}
\mbox{~\epsfig{file=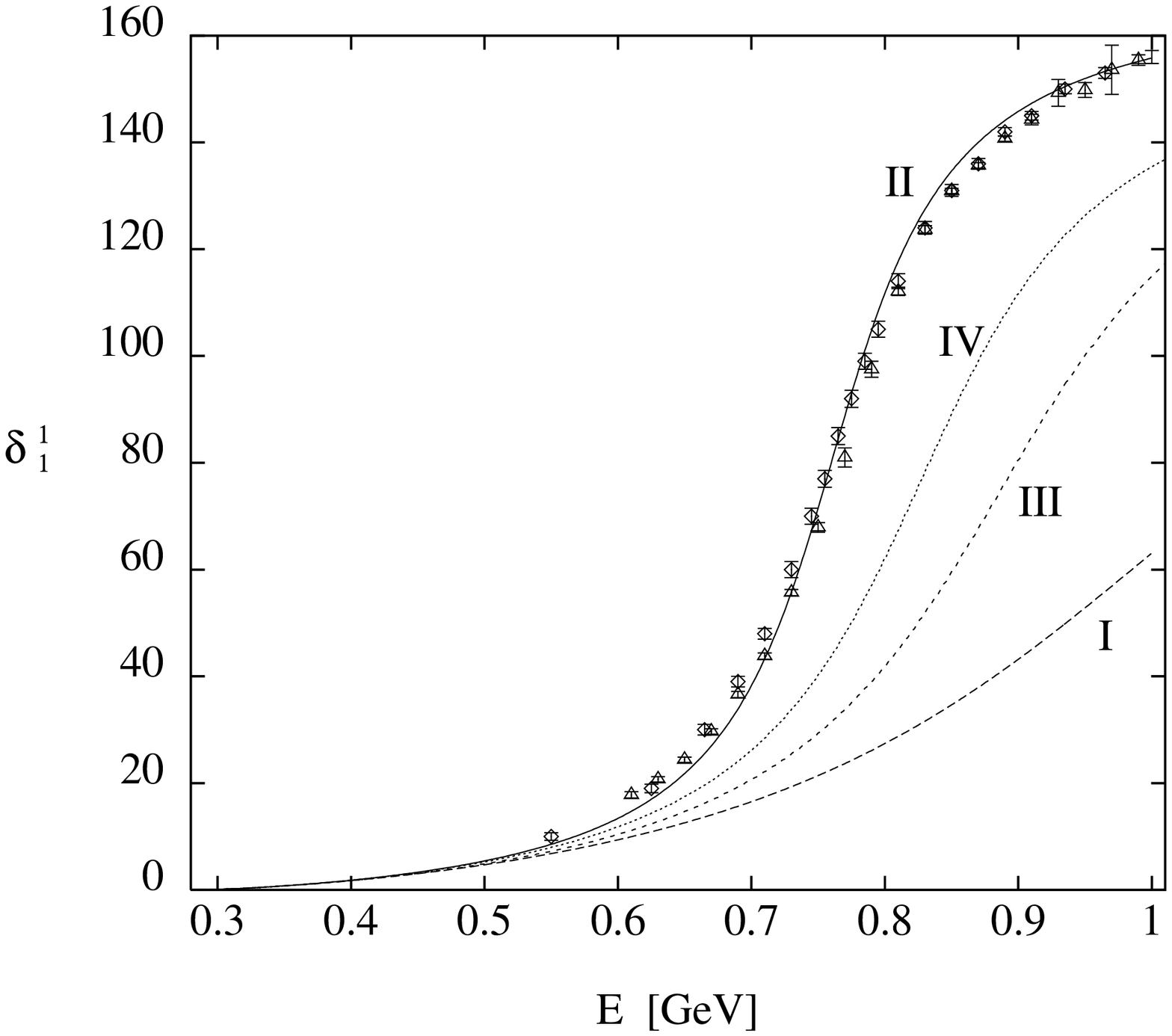,angle=-90,width=15cm}}
\caption{\leftskip = 1cm 
The $\pi \pi$ $I=1$ $P$-wave phase shift, $\delta ^1 _1$ in degrees, below 
$K \overline K$ threshold as a function of $\pi\pi$ mass 
$E\,=\, \protect \sqrt{s}$. 
The dashed line  is the result
obtained with no left hand cut contribution (Scheme~I); the solid 
line comes from the explicit evaluation of the left hand cut dispersive 
integral (Scheme~II); the dotted lines are obtained with additional
summation assumptions (Schemes~III and IV). The experimental data are from 
Protopopescu et al.~[10] (diamonds) and the energy-independent analysis by
Ochs~[11] (triangles).}
\end{center}
\end{figure}
\newpage
\begin{figure}[p]
\begin{center}
\mbox{~\epsfig{file=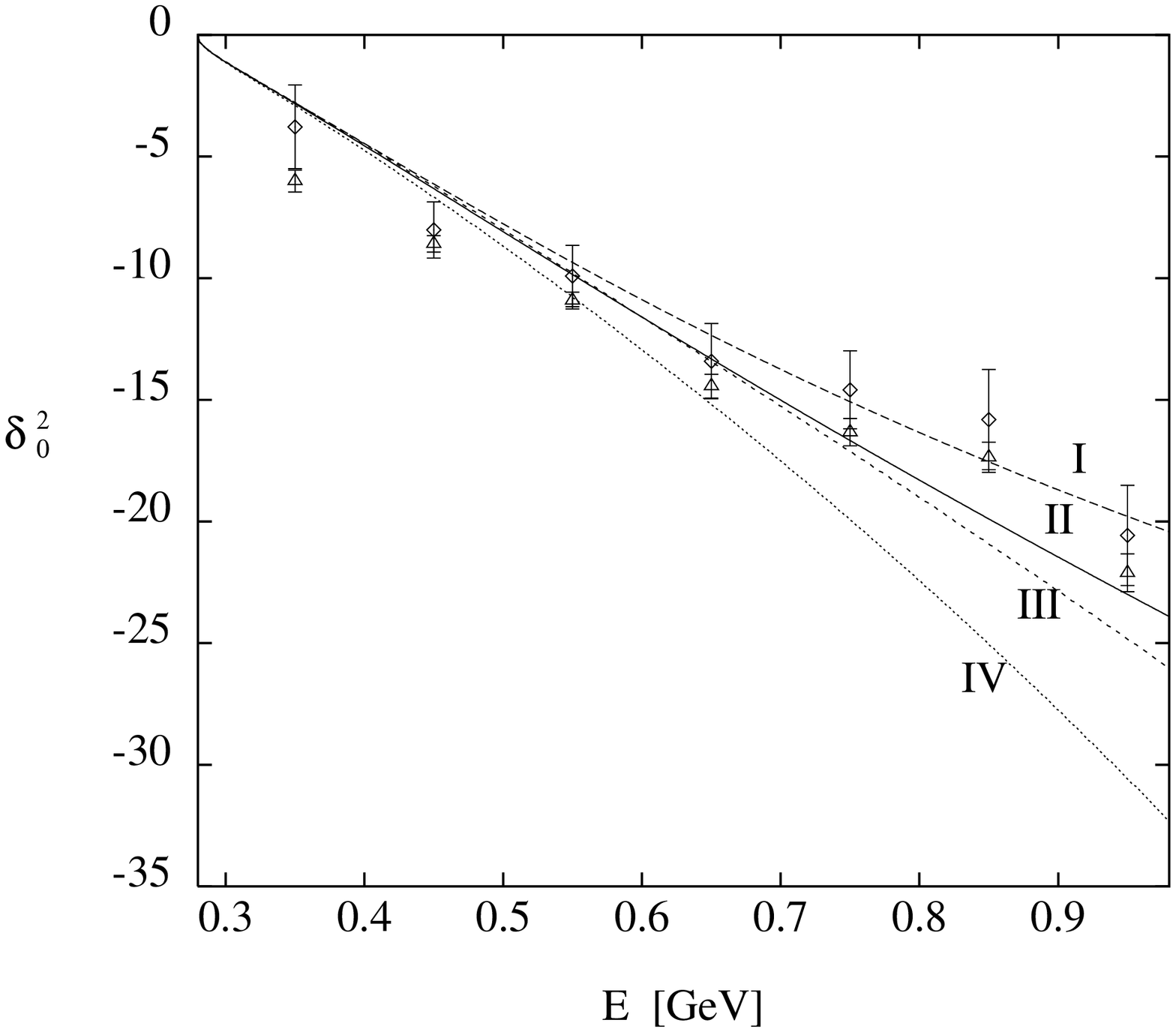,angle=-90,width=15cm}}
\caption{\leftskip = 2cm 
The $\pi \pi$ $I=2$ $S$-wave phase shift, $\delta ^2 _2$ in degrees, below 
$K \overline K$ threshold as a function of $\pi\pi$ mass 
$E\,=\, \protect \sqrt{s}$.
 The dashed line  is the result
obtained with no left hand cut contribution (Scheme~I); the solid 
line comes from the explicit evaluation of the left hand cut dispersive 
integral (Scheme~II); the dotted lines are obtained with additional
summation assumptions (Schemes~III and IV). 
The experimental data are from the analyses by Hoogland et al.~[12].}
\end{center}
\end{figure}
\end{document}